\documentclass[aps,onecolumn]{revtex4}%
\usepackage{amsfonts}
\usepackage{amsmath}
\usepackage{amssymb}
\usepackage{graphicx}%
\begin{document}
\title{Radiation Entropy in asymptotically AdS Black Holes within $f(\mathbb{Q})$ Gravity}
\author{Yipeng Liu}
\affiliation{School of Mathematics and Physics, China University of Geosciences, Wuhan
430074, China}
\author{Wei Xu}
\affiliation{School of Mathematics and Physics, China University of Geosciences, Wuhan
430074, China}
\author{Baocheng Zhang}
\email{zhangbaocheng@cug.edu.cn}
\affiliation{School of Mathematics and Physics, China University of Geosciences, Wuhan
430074, China}
\begin{abstract}
We employ the island rule to investigate the radiation entropy of
asymptotically AdS black holes in the framework of $f(\mathbb{Q})$ gravity. 
Through an analysis based on the Euclidean action, we find that
 the area term of the generalized entropy must be modified, 
which in turn leads to a modification of the island rule itself. 
Using the corrected rule to compute the radiation entropy for the eternal 
asymptotically AdS black hole reveals that, 
the result diverges as the cutoff surface is moved outward, 
indicating the breakdown of the $s$-wave approximation. 
For a collapsing asymptotically AdS black hole, 
the radiation entropy is dominated by the area term, 
with a logarithmic correction proportional to the area,
which is consistent with the predictions of quantum gravity theories.
Furthermore, both the radiation entropy and the Page time are ultimately
influenced by the choice of the $f(\mathbb{Q})$ model, 
implying that information regarding the underlying gravitational model 
is encoded in the final radiation entropy.

\end{abstract}
\maketitle


\section{Introduction}

Since the discovery of Hawking radiation \cite{Hawking:1974rv,Hawking:1975vcx}%
, the fundamental nature of the emitted radiation remains unclear.
Hawking's original semi-classical calculation suggests that 
the final state of black hole evaporation is a mixed state, 
independent of the initial state that formed the black hole. 
This apparent violation of unitarity is known as
the black hole information loss paradox \cite{Hawking:1976ra}. 
A complete resolution of this paradox requires a full theory of quantum gravity, 
which, however, has not yet been established.

On the other hand, as a low-energy effective theory of quantum gravity,
general relativity itself also calls for modification. 
For instance, to account for the accelerated expansion of the universe
\cite{SupernovaSearchTeam:1998fmf,SupernovaCosmologyProject:1998vns,Boomerang:2000efg,Hanany:2000qf,Riess:2011yx,Planck:2015fie,Planck:2018nkj,Planck:2018vyg}%
, one has to introduce dark energy that constitute about two-thirds of the
total cosmic content, yet it has not been directly detected so far.
Investigating the information loss problem in the context of modified gravity,
therefore, can provide further insights into quantum gravity.

As an extension of the symmetric teleparallel equivalent of general
relativity, $f(\mathbb{Q})$ gravity has attracted considerable attention
recently
\cite{BeltranJimenez:2017tkd,BeltranJimenez:2018vdo,BeltranJimenez:2019tme}.
Current results indicate that this theory can challenge the cold dark matter
model without the need for dark components \cite{Shi:2023kvu, Atayde:2021pgb,
Anagnostopoulos:2021ydo}. In addition, the theory itself exhibits several
notable advantages. For example, compared with gravitational theories based on
the Riemannian geometry, $f(\mathbb{Q})$ gravity possesses a well-defined
variational principle without the need for additional boundary terms
\cite{Heisenberg:2023lru}. This thus provides an improved theoretical framework
for investigating the information paradox.

In general relativity, the key to resolving this paradox lies in computing the
entanglement entropy of a gravitational system. Progress has been made through
the holography principle
\cite{Susskind:1994vu,Maldacena:1997re,Ryu:2006bv,Faulkner:2013ana}, which
leads to a prescription for computing the entanglement entropy of
gravitational systems, known as the island rule
\cite{Engelhardt:2014gca,Penington:2019npb,Almheiri:2019psf,Almheiri:2019hni}%
:
\begin{equation}
S_{R}=Min_{X}\left\{  Ext_{X}\left[  \frac{\mathcal{A}(X)}{G_{N}}%
+S_{se-cl}(\Sigma_{R}\cup\Sigma_{I})\right]  \right\}  , \label{ir1}%
\end{equation}
where $X$ is referred to as the quantum extreme surface, $\mathcal{A}$ represents its
area, and $S_{se-cl}(\Sigma_{R}\cup\Sigma_{I})$ denotes the semi-classical
entropy of the matter fields on both the radiation and island regions. The
entire expression inside the square brackets is referred to as the generalized
entropy, whose first term is proportional to the horizon area and originates
from the Ryu-Takayanagi (RT) formula \cite{Ryu:2006bv}.

This rule was later justified using the replica trick and gravitational path
integrals
\cite{Lewkowycz:2013nqa,Dong:2017xht,Penington:2019kki,Almheiri:2019qdq,Dong:2016hjy}%
, and it has since been generalized to arbitrary spacetimes beyond AdS
\cite{Almheiri:2020cfm}. Recently, the island rule has attracted considerable
attention due to its ability to reproduce the Page curve
\cite{Page:1993wv}. Most existing studies are based on gravitational
theories formulated within the framework of Riemannian geometry, such as
the two-dimensional Jackiw-Teitelboim gravity \cite{Almheiri:2019qdq}, eternal
Schwarzschild black holes \cite{Hashimoto:2020cas}, charged
Reissner-Nordstr\"{o}m black holes \cite{Wang:2021woy}, among others
\cite{Arefeva:2021kfx,Du:2022vvg,Anand:2022mla,Wang:2024itz,Yu:2024fks,Liu:2025flo}%
. It is generally found that the radiation entropy is time-independent at the late
stage of evaporation, with the corresponding quantum extremal surface (or
island) lying outside the horizon \cite{Almheiri:2019yqk,He:2021mst}.

Nevertheless, some issues still remain here. First, as candidates for
the quantum gravity theory, both the string theory and loop quantum gravity suggest that
the black hole entropy receives a logarithmic correction proportional to the event horizon area
\cite{Fursaev:1994te,Kaul:2000kf,Page:2004xp,Zhang:2008rd,Ghosh:2004rq,Solodukhin:1997yy,Xiang:2006mg}%
, when quantum effects are considered. According to the properties of
entanglement entropy, for a black hole formed from a pure state, the radiation
entropy should coincide with the black hole entropy \cite{Srednicki:1993im}.
However, in the existing literature on the island rule, such a result has not
been observed. Furthermore, the situation becomes more problematic in the
case of the eternal black hole. Under the s-wave approximation, the radiation
entropy diverges as the cutoff surface is moved farther from the horizon.
Since the cutoff surface is fictitious and the region beyond it is assumed to
be weakly gravitating, this result indicates that the s-wave approximation fails for the eternal
black hole. Considering a collapsing black hole can avoid the issues
associated with the s-wave approximation \cite{Gan:2022jay}, but the
divergence of the radiation entropy with respect to the cutoff surface cannot
be avoided if the island is located inside the horizon.

In this work, we study the information paradox and the island rule in the context
of $f(\mathbb{Q})$ gravity. In Sec.~\ref{secII}, we introduce the fundamentals of
$f(\mathbb{Q})$ gravity, including its action and field equations, and present
black hole solutions together with the generalized entropy. In Sec.~\ref{secIII}, we
evaluate the radiation entropy of the eternal black hole, address the
divergence issue, and then extend the analysis to a collapsing black hole background.
Finally, a summary of the main results is provided in Sec.~\ref{secIV}.

\section{General Entropy in $f(\mathbb{Q})$ Gravity}
\label{secII}

\subsection{Elements}

In $f(\mathbb{Q})$ gravity, the metric $g_{\mu\nu}$ and the affine connection
$\Gamma^{\alpha}{}_{\mu\nu}$ are two independent variables. The geometry is
described by the non-metricity tensor,
\begin{equation}
Q_{\alpha\mu\nu}=\nabla_{\alpha}g_{\mu\nu}=\partial_{\alpha}g_{\mu\nu}%
-2\Gamma^{\lambda}{}_{\alpha(\mu}g_{\nu)\lambda},
\end{equation}
but not the Riemann tensor. From this tensor one can construct the non-metricity
scalar, with its form being
\begin{equation}
\mathbb{Q}=\frac14 Q_{\alpha\beta\gamma}Q^{\alpha\beta\gamma}-\frac12
Q_{\alpha\beta\gamma}Q^{\beta\alpha\gamma}-\frac14 Q_{\alpha}Q^{\alpha}%
+\frac12 Q_{\alpha}\tilde{Q}^{\alpha},
\end{equation}
where $Q_{\alpha}=Q_{\alpha}{}^{\mu}{}_{\mu}$ and $\tilde{Q}_{\alpha}=Q^{\mu
}{}_{\alpha\mu}$ are respectively different traces of the non-metricity tensor.

The action of $f(\mathbb{Q})$ theory is given by \cite{Heisenberg:2023lru}
\begin{equation}
I=-\frac{1}{2 k} \int_{\mathcal{M}} d^{4}x \sqrt{-g} f(\mathbb{Q}) +I_{m},
\end{equation}
where $k=8\pi G_{N}$ is the gravitational constant, $f$ is an arbitrary
function of the non-metricity scalar, and $I_{m}$ is the action of matter.
Variations of this action with respect to the metric and the connection yield
\cite{Heisenberg:2023lru}
\begin{align}
\mathcal{E}_{\mu\nu}  &  \equiv\frac{2}{\sqrt{-g}}\nabla_{\alpha}(\sqrt
{-g}f_{\mathbb{Q}}P^{\alpha}{}_{\mu\nu})-\frac12g_{\mu\nu}f+f_{\mathbb{Q}}
q_{\mu\nu} =k T_{\mu\nu},\label{eeq}\\
\mathcal{C}_{\alpha}  &  \equiv\nabla_{\mu}\nabla_{\nu}(\sqrt{-g}
f_{\mathbb{Q}} P^{\mu\nu}{}_{\alpha})=0. \label{peq}%
\end{align}
For convenience, the entire paper uses the notation $f_{\mathbb{Q}}\equiv
df/d\mathbb{Q}$ and $f_{\mathbb{Q} \mathbb{Q}} \equiv d^{2}f/d\mathbb{Q}^{2}$.
The non-metricity conjugate and the symmetric tensor are defined as
\begin{align}
P^{\alpha}{}_{\mu\nu} &=-\frac14 Q^{\alpha}{}_{\mu\nu}+\frac12 Q_{(\mu}%
{}^{\alpha}{}_{\nu)}+\frac14(Q^{\alpha}-\tilde{Q}^{\alpha})g_{\mu\nu}%
-\frac14\delta^{\alpha}{}_{(\mu}Q_{\nu)},\\
q_{\mu\nu} &=P_{(\mu|\alpha\beta|} Q_{\nu)}{}^{\alpha\beta}-2Q_{ \alpha
\beta(\mu} P^{\alpha\beta}{}_{\nu)}.
\end{align}
The energy-momentum tensor, which is the same as that in general relativity, 
is given by
\begin{equation}
T_{\mu\nu}=-\frac{2}{\sqrt{-g}}\frac{\delta I_{m}}{\delta g^{\mu\nu}},
\end{equation}

Setting $f(\mathbb{Q})=\mathbb{Q}-2\lambda$, the metric field equation will
return to the symmetric teleparallel equivalent of general relativity,
which is dynamically equivalent to general relativity itself. To clarify this point,
 the field equation can be rewritten in a more compact form,
\begin{equation}
f_{\mathbb{Q}}\mathcal{G}_{\mu\nu}-\frac{1}{2}(f-f_{\mathbb{Q}}\mathbb{Q}%
)g_{\mu\nu}+2f_{\mathbb{Q}\mathbb{Q}}P^{\alpha}{}_{\mu\nu}\partial_{\alpha
}\mathbb{Q}=kT_{\mu\nu}, \label{feGR}%
\end{equation}
where $\mathcal{G}_{\mu\nu}$ is the Einstein tensor. Clearly, the above equation can be reduced
to the Einstein equation for this special choice. On the other hand, from the
connection field equation one can obtain \cite{Heisenberg:2023lru}
\begin{equation}
\Gamma^{\alpha}{}_{\mu\nu}=\frac{\partial x^{\alpha}}{\partial\xi^{\lambda}%
}\partial_{\mu}\partial_{\nu}\xi^{\lambda},
\end{equation}
where $\xi^{\lambda}$ represents four arbitrary functions of coordinates. If we
choose $x^{\mu}=\xi^{\mu}$, the connection then vanishes globally, which is
known as the coincident gauge \cite{BeltranJimenez:2017tkd}.

\subsection{Black holes}

Choosing such a metric ansatz
\begin{equation}
ds^{2}=-a(r)dt^{2}+\frac{1}{a(r)}dr^{2}+r^{2}(dx^{2}+dy^{2}),
\end{equation}
where the event horizon has a flat spatial topology, the corresponding
non-metricity scalar is
\begin{equation}
\mathbb{Q}=\frac{2a}{r}\left(  \frac{1}{r}+\frac{a^{\prime}}{a}\right)  .
\label{nms}%
\end{equation}
Considering a vacuum, the non-trivial components of the metric field equation are
\begin{align}
0  &  =\frac{f}{2}-f_{\mathbb{Q}}\mathbb{Q}-\frac{2a}{r}f_{\mathbb{Q}%
\mathbb{Q}}\mathbb{Q}^{\prime},\\
0  &  =-\frac{f}{2}+f_{\mathbb{Q}}\mathbb{Q},\\
0  &  =-\frac{f}{2}+f_{\mathbb{Q}}\left(  \mathbb{Q}+\frac{r}{4}%
\mathbb{Q}^{\prime}\right)  +f_{\mathbb{Q}\mathbb{Q}}\mathbb{Q}^{\prime
}\left(  \frac{\mathbb{Q}r}{2}-\frac{a^{\prime}}{2}\right)  ,
\end{align}
The above three equations are not independent; after simplification, only two
valid field equations remain,
\begin{align}
0  &  =-\frac{f}{2}+f_{\mathbb{Q}}\mathbb{Q},\label{fQM}\\
0  &  =-\frac{2a}{r}f_{\mathbb{Q}\mathbb{Q}}\mathbb{Q}^{\prime}.
\end{align}
The first equation merely imposes a constraint on $f$, 
while the second is of real utility, from which we obtain
\begin{equation}
f_{\mathbb{Q}\mathbb{Q}}=0\quad\;\text{or}\;\quad\mathbb{Q}^{\prime}=0.
\end{equation}
Since $f_{\mathbb{Q}\mathbb{Q}}=0$ returns to general relativity, our main interest 
then focuses on $\mathbb{Q}^{\prime}=0$, which means the
non-metricity scalar is a constant, denoted by $\mathbb{Q}_{0}$. Subsequently,
by directly solving Eq.~(\ref{nms}), we obtain
\begin{equation}
a(r)=\frac{r^{2}}{\ell^{2}}-\frac{2m}{r},
\end{equation}
and the final metric becomes
\begin{equation}
ds^{2}=-\left(  \frac{r^{2}}{\ell^{2}}-\frac{2m}{r}\right)  dt^{2}+\left(
\frac{r^{2}}{\ell^{2}}-\frac{2m}{r}\right)  ^{-1}dr^{2}+r^{2}(dx^{2}+dy^{2}),
\label{bh}%
\end{equation}
where $m$ is the black hole mass, and $\ell$ is related to the cosmological
constant, which satisfies $\mathbb{Q}_{0}\ell^{2}=6$. To ensure the existence of an event 
horizon, the non-metricity scalar must take positive values. The metric then
denotes an asymptotically AdS black hole with a flat horizon
\cite{Cvetic:1999xp}.

It is unsurprising that this solution also belongs to general relativity,
since for a constant non-metricity scalar Eq.~(\ref{feGR}) becomes
\begin{equation}
\mathcal{G}_{\mu\nu}-\frac{\mathbb{Q}_{0}}{2}g_{\mu\nu}=\frac{k}%
{f_{\mathbb{Q}}}T_{\mu\nu},
\end{equation}
which can be regarded as the Einstein equation with an effect cosmological
constant and a rescaling energy-momentum tensor. When an electromagnetic field
exists, the $f(\mathbb{Q})$ gravity yields charged black hole solutions that
go beyond those of general relativity \cite{Nashed:2023tua}.
Setting the charge to zero recovers the uncharged solution obtained here.
But due to the difference in the action, this black hole exhibits 
certain thermodynamic properties distinct from those in general relativity. 
In particular, the form of the generalized entropy is modified. Consequently, the
island rule, formulated based on the generalized entropy to address
the information loss problem, needs to be reconsidered within the framework of
$f(\mathbb{Q})$ gravity.

\subsection{Generalized Entropy}

When quantum effects are taken into account, the black hole acquires a Hawking
temperature given by
\begin{equation}
T=\frac{\kappa}{2\pi}=\frac{3r_{h}}{4\pi\ell^{2}}\equiv\beta^{-1},
\end{equation}
where $\kappa$ is the surface gravity, $r_{h}=\sqrt[3]{(2m\ell^{2})}$ is the event
horizon, and $\beta$ is the imaginary period. The partition function for a
gravitational system with quantum field is \cite{Almheiri:2020cfm}
\begin{equation}
\mathcal{Z}\sim e^{I_{E}}\mathcal{Z}_{quantum},
\end{equation}
where $I_{E}$ is the classical Euclidean action, and $\mathcal{Z}_{quantum}$
is related to the quantum fields. The generalized entropy can then be derived as
\begin{align}
S_{gen}=  &  (1-\beta\partial_{\beta})\ln\mathcal{Z}\nonumber\label{sgen}\\
=  &  \frac{2f_{\mathbb{Q}}\pi r_{h}^{2}}{G_{N}}+S_{se-cl}(\Sigma),
\end{align}
where we have used the classical Euclidean action $I_E$, which is given by
\begin{align}
I_{E}  &  =\lim_{R\rightarrow\infty}\left(  \frac{V(\mathcal{M}^{2})}{16\pi
G_{N}}\int_{0}^{\beta}dt\int_{r_{h}}^{R}dr[f(\mathbb{Q}_{0})r^{2}]\right.
\nonumber\\
&  \left.  -\frac{V(\mathcal{M}^{2})}{16\pi G_{N}}\int_{0}^{\beta^{\prime}%
}dt\int_{0}^{R}dr[f(\mathbb{Q}_{0})r^{2}]\right) \nonumber\\
&  =-\frac{2f_{\mathbb{Q}}\pi r^{2}}{3G_{N}}%
\end{align}
with $\beta$ and $\beta^{\prime}$ satisfying
\cite{Gibbons:1976ue,Hawking:1982dh,Birmingham:1998nr}
\begin{equation}
\beta^{\prime}\sqrt{\frac{R^{2}}{\ell^{2}}}=\beta\sqrt{\frac{R^{2}}{\ell^{2}%
}-\frac{2m}{R}}.
\end{equation}
It should be noted that the generalized entropy of black holes depends on the
location of their event horizon. Unlike in general relativity, however, the area term
now carries a coefficient $f_{\mathbb{Q}}$, implying that the specific
$f(\mathbb{Q})$ gravity model will affect the black hole entropy. Such an
influence is naturally expected to manifest in the radiation entropy as well.
Furthermore, a similar type of modification also appears in $f(\mathcal{R})$
gravity \cite{Cognola:2011nj,Zheng:2018fyn}. But, unlike the
$f(\mathcal{R})$ case, in $f(\mathbb{Q})$ theory this form holds only for
vacuum solutions; the situation in the presence of matter fields requires 
further careful investigation.

Finally, to recover general relativity in $f(\mathbb{Q})$ gravity, 
one typically needs to set $f_{\mathbb{Q}}=1$ \cite{DAmbrosio:2021zpm,DAmbrosio:2021pnd}%
. This condition implies that the radiation entropy in $f(\mathbb{Q})$ gravity
is twice that of general relativity. The discrepancy originates from the
difference in boundary terms between the two theories. Although such terms do
not affect the field equations, they do influence the entanglement entropy
once quantum effects are taken into account. Since $f(\mathbb{Q})$ gravity
does not require the artificial introduction of additional boundary terms, this
may suggest that the boundary terms in general relativity itself should be
reconsidered or modified.

\section{Entropy of the Hawking radiation}
\label{secIII}

\subsection{Island Rule}

While the form of the island rule is given in Eq.~(\ref{ir1}), in $f(\mathbb{Q})$
gravity, the rule should take the following formula:
\begin{equation} \label{ir2}
S_{R}=Min_{X}\left\{  Ext_{X} \left[  \frac{f_{\mathbb{Q}} \mathcal{A}%
(X)}{2G_{N}} +S_{se-cl}(\Sigma_{R}\cup\Sigma_{I}) \right]  \right\}  .
\end{equation}
Here we provide an explanation as follows:

\begin{itemize}
\item The entire expression within the brackets still represents the
generalized entropy, but it now depends on the quantum extremal surface $X$. The
first term of the generalized entropy is the area term, originally proposed by
the RT formula, which is derived from the holographic principle and inspired by
the area law of black hole entropy. Unlike in general relativity, in
$f(\mathbb{Q})$ gravity the black hole entropy depends not only on the horizon
area but also on the specific form of the $f(\mathbb{Q})$ model. Consequently,
one can infer that the RT formula itself will also be modified by this model,
such as
\begin{equation}
S=\frac{f_{\mathbb{Q}} \mathcal{A}_{M}}{2G_{N}},
\end{equation}
where $\mathcal{A}_{M}$ denotes the area of minimal surface \cite{Ryu:2006bv}.
The generalized form then becomes the same as Eq.~(\ref{sgen})
\cite{Faulkner:2013ana}. Subsequently, considering the effect of the quantum fields
on the minimal surface \cite{Engelhardt:2014gca}, the final radiation entropy
takes the formula of Eq.~(\ref{ir2}).

\item The second term of generalized entropy is referred to as the semi-classical
entropy, which essentially represents the entanglement entropy of quantum
fields in curved spacetime. Although $f(\mathbb{Q})$ gravity employs a
different geometric quantity to describe gravitation, the nature of spacetime
itself remains unchanged, as the theory is dynamically equivalent to the Riemannian
description. Therefore, the form of the semi-classical entropy here remains
the same as in general relativity. However, the $f(\mathbb{Q})$ model can
modify the properties of black hole solutions, such as the AdS radius, thereby
influencing the semi-classical contribution. Moreover, the procedure for
computing the semi-classical entropy can still follow the standard results in
general relativity. For instance, when the cutoff surface is far from the
event horizon, the semi-classical entropy can be evaluated using the s-wave
approximation \cite{Calabrese:2004eu,Calabrese:2009qy},
\begin{equation}
S_{se-cl}[ds^{2}]=\frac{c}{3}\ln d(I,A),
\end{equation}
where $c$ is the central charge, and $d(I,A)$ is the distance between $I$ and
$A$. Under the conformal transformation, the entanglement entropy becomes
\cite{Almheiri:2019psf,Gan:2022jay},
\begin{equation}
S_{se-cl}[\Omega^{2}ds^{2}]=\frac{c}{6}\ln[d(I,A)^{2}\Omega(A)\Omega(I)],
\label{sw}%
\end{equation}
with $d(I,A)$ still being calculated in the previous metric $ds^{2}$.

\item 
The initial state considered in this paper is pure. According to the
properties of entanglement entropy, there are
\begin{equation}
	S_{se-cl}(\Sigma_{I}\cup\Sigma_{R}) =S_{se-cl}(\Sigma_{X}),
\end{equation}
where $\Sigma_{X}$ is the region between the quantum extremal surface and
the cutoff surface, which is referred to as the black hole region. 
By this condition, the island rule can then be rewritten as
\begin{equation}
	S_{R}=Min_{X}\left\{  Ext_{X} \left[  \frac{f_{\mathbb{Q}} \mathcal{A}%
		(X)}{2G_{N}} +S_{se-cl}(\Sigma_{X}) \right]  \right\}  .
\end{equation}
The radiation entropy will be calculated based on this formula.

\end{itemize}

\subsection{Eternal Scenarios}

\begin{figure}[ptb]
\centering
\includegraphics[width=0.4\linewidth]{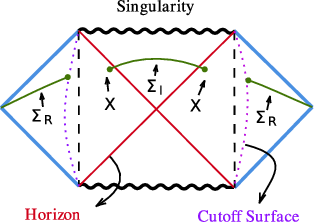} \caption{The island
configuration in an eternal AdS black hole, where $X$ denotes the quantum
extremal surface. The left and right regions in the figure are symmetric.}%
\label{Fig1}%
\end{figure}

When considering the information paradox, it is convenient for us to 
fix the angular coordinates, $dx=dy=0$. 
Under Kruskal coordinates transformation,
$U=-e^{-\kappa(t-r^{\ast})},\;V=e^{\kappa(t+r^{\ast})}$, the metric then becomes
\begin{equation}
ds^{2}=-\Omega^{2}dUdV=-\frac{a(r)}{\kappa^{2}}e^{-2\kappa r^{\ast}}dUdV,
\end{equation}
where $r^{\ast}$ is the tortoise coordinate, with $dr^{\ast}/dr=1/a(r)$. Since an
asymptotic AdS black hole itself cannot completely evaporate, a thermal bath
coupled to radiation is required, and the cutoff surface is then assumed to be 
located in it. 
A detailed schematic illustration can be found in Fig.~\ref{Fig1}.
The conformal factor for the thermal bath is
\begin{equation}
\Omega^{2}=\frac{1}{\kappa^{2}}e^{-2\kappa r^{\ast}}.
\end{equation}

\begin{figure}[ptb]
\centering
\includegraphics[width=0.4\linewidth]{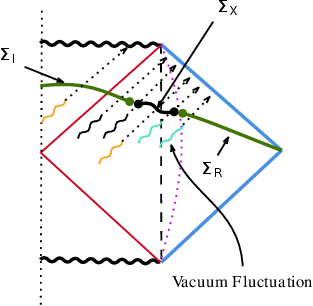} \caption{The vacuum
fluctuations near the island and the cutoff surface, and only the right-half
spacetime structure of the eternal black hole is illustrated here.}%
\label{Fig2}%
\end{figure}

For convenience, we denote the coordinates of the island and the cutoff
surface as $(t_{I},r_{I})$ and $(t_{A},r_{A})$, respectively. Then using the
Kruskal coordinates, the generalized entropy becomes
\begin{align}
S_{gen}  &  =f_{\mathbb{Q}}\frac{4\pi r_{I}^{2}}{G_{N}}+\frac{c}{6}\ln\left[
(U_{I}-U_{A})^{2}(V_{A}-V_{I})^{2}\Omega^{2}(A)\Omega^{2}(I)\right]
\nonumber\\
&  =f_{\mathbb{Q}}\frac{4\pi r_{I}^{2}}{G_{N}}+\frac{c}{6}\ln\left[
\frac{4a(r_{I})}{\kappa^{4}}\right. \nonumber\\
&  \left.  \times(\cosh[\kappa(r_{A}^{\ast}-r_{I}^{\ast})]-\cosh[\kappa
(t_{A}-t_{I})])^{2}\right]  .
\end{align}
Firstly, the derivative of $S_{gen}$ with respect to $t_{I}$ yields
\begin{equation}
\frac{\partial S_{gen}}{\partial t_{I}}=\frac{c\kappa\sinh[\kappa(t_{A}%
-t_{I})]}{3(\cosh[\kappa(r_{A}^{\ast}-r_{I}^{\ast})]-\cosh[\kappa(t_{A}%
-t_{I})])},
\end{equation}
which gives
\begin{equation}
t_{I}=t_{A}. \label{til}%
\end{equation}
Submitting this condition into $S_{gen}$ and then taking its derivative with
respect to $r_{I}$, we obtain
\begin{align}
\frac{\partial S_{gen}}{\partial r_{I}}  &  =f_{\mathbb{Q}}\frac{8\pi r_{I}%
}{G_{N}}+\frac{ca^{\prime}(r_{I})}{6a(r_{I})}-\frac{2c\kappa}{6a(r_{I})}%
\coth\left[  \frac{\kappa}{2}(r_{A}^{\ast}-r_{I}^{\ast})\right] \nonumber\\
&  =f_{\mathbb{Q}}\frac{8\pi r_{I}}{G_{N}}+\frac{c}{6a(r_{I})}(a^{\prime
}(r_{I})-2\kappa)-\frac{2c\kappa}{3a(r_{I})}e^{-\kappa(r_{A}^{\ast}%
-r_{I}^{\ast})}\nonumber\\
&  =f_{\mathbb{Q}}\frac{8\pi r_{I}}{G_{N}}-\frac{2c\kappa}{3a(r_{I}%
)}e^{-\kappa(r_{A}^{\ast}-r_{I}^{\ast})},
\end{align}
where we have used $\coth y\simeq1+2e^{-2y}\;(y\rightarrow+\infty)$ and
$a^{\prime}(r_{I})-2\kappa\simeq a^{\prime\prime}(r_{I})(r_{I}-r_{h}%
)^{2}\simeq0$. Setting $r_{I}=r_{h}+x^{2}r_{h}$ and assuming $r_{A}\gg
r_{h}\gg x$, the above equation can be simplified as
\begin{equation}
f_{\mathbb{Q}}\frac{8\pi r_{I}}{G_{N}}-\frac{2c\kappa}{3xr_{h}}e^{-\kappa
r_{A}^{\ast}}\simeq0,
\end{equation}
which gives
\begin{equation}
x=\frac{1}{f_{\mathbb{Q}}}\frac{cG_{N}\kappa}{12\pi r_{h}}e^{-\kappa
r_{A}^{\ast}}.
\end{equation}
The location of the island then is
\begin{equation}
r_{I}=r_{h}+\frac{1}{f_{\mathbb{Q}}^{2}}\frac{(cG_{N}\kappa)^{2}}{144\pi
^{2}r_{h}}e^{-2\kappa r_{A}^{\ast}}. \label{ril}%
\end{equation}
Finally, submitting Eqs.~(\ref{til}) and (\ref{ril}) into $S_{gen}$, we find that 
the radiation entropy takes the following form,
\begin{align}
S_{R}  &  \simeq f_{\mathbb{Q}}\frac{4\pi r_{I}^{2}}{G_{N}}+\frac{c}{6}%
\ln\left[  \frac{a(r_{I})}{\kappa^{4}}e^{2\kappa(r_{A}^{\ast}-r_{I}^{\ast}%
)}\right] \nonumber\\
&  \simeq f_{\mathbb{Q}}\frac{4\pi r_{h}^{2}}{G_{N}}+\frac{c}{6}\ln\left[
\frac{2}{\kappa^{3}}e^{2\kappa r_{A}^{\ast}}\right]  .
\end{align}
We provide a discussion of this result as follows:

\begin{itemize}
\item[(1)] From the final expression, we can see that the radiation
entropy no longer evolves with time. What causes this behavior, and what does
it imply? In quantum theory, the vacuum is not empty but filled with quantum
fluctuations, and Hawking radiation itself originates from such fluctuations
near the horizon. Now, consider the vacuum fluctuations occurring near the
quantum extremal surface and the cutoff surface, as shown in Fig.~\ref{Fig2}, which illustrates
only the right half of the eternal AdS black hole.

To obtain a finite radiation entropy, namely a finite-dimensional quantum
Hilbert space, these vacuum fluctuations must inevitably lead to the absence
of some of the original Hawking quanta in the black hole region, which is denoted by
$\Sigma_{X}$. For instance, a pair fluctuation near the quantum extremal
surface (shown as black curves in Fig.~2) can cause an early Hawking particle
that had fallen into the black hole to escape into the radiation region,
thereby restoring information. As a result, for an eternal black hole, in
order to recover a certain amount of information, an equal amount of
information must necessarily be lost.

\item[(2)] Another important feature is that the radiation entropy depends
on the $f(\mathbb{Q})$ model. This dependence appears not only
in the coefficient in front of the area term, but also in the
horizon radius and the surface gravity. Referring back to Eq.~(\ref{fQM}), one
can see that different $f(\mathbb{Q})$ models lead to different non-metricity
scalars (and thus different AdS radii), which in turn result in distinct
horizon radii and surface gravities. Of course, these effects will likewise
influence the location of the quantum extremal surface.

\item[(3)] It should be noted that this radiation entropy suffers from a
significant problem. When the cutoff surface is moved further away from the horizon,
the result grows without bound and eventually diverges. This behavior is
inconsistent, first of all, with the $s$-wave approximation itself, which
inherently assumes that the cutoff surface is far from the horizon. For a
sufficiently distant cutoff surface, the area term can no longer remain the
dominant contribution to the radiation entropy.

Secondly, this divergent result also contradicts the physical picture. For
instance, in Fig.~2, if we approximately set $\Sigma_{R}\simeq0$, the entire
exterior region of the horizon can be regarded as $\Sigma_{X}$, which together
with $\Sigma_{I}$ forms a pure state. From Eq.~(\ref{ril}) it is evident that
the size of the island region $\Sigma_{I}$ remains essentially unchanged,
since the boundary of the island approaches the horizon infinitely as the
cutoff surface moves farther away. A divergent entropy in this case would thus
imply that $\Sigma_{I}$, being a finite region, possesses an infinite Hilbert space.

Considering this, the most plausible explanation is that the $s$-wave
approximation does not hold in the case of an eternal black hole, since
maintaining the black hole mass necessarily requires the presence of both
ingoing and outgoing modes at spatial infinity \cite{Gan:2022jay}.
\end{itemize}

\subsection{Collapsing Scenarios}

By replacing the mass function with a step function model, we can otatin 
a simple collapsed black hole model, whose corresponding metric is given by
\begin{equation}
ds^{2}=a(v,r)dv^{2}+2dvdr.
\end{equation}
When $v<v_{0}$ the metric describes a conformal flat spacetime, $a(v,r)={r^{2}%
}/{\ell^{2}}\equiv a_{0}(r)$; otherwise, it is an AdS black hole,
$a(v,r)={r^{2}}/{\ell^{2}}-{2m}/{r}\equiv a(r)$.

\begin{figure}[ptb]
\centering
\includegraphics[width=0.4\linewidth]{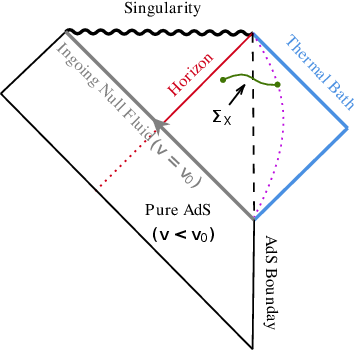} \caption{A black hole formed
by the collapse of an AdS vacuum, with its right exterior region coupled to a
thermal bath that collects the Hawking radiation.}%
\label{Fig3}%
\end{figure}

In double-null coordinates, the metric becomes
\begin{equation}
ds^{2}=%
\begin{cases}
-a_{0}(r)dudv, & v<v_{0}\\
-a(r)d\tilde{u}dv, & v\geq v_{0}%
\end{cases}
,
\end{equation}
where $u=v+2\ell^{2}/r,\;\tilde{u}=v-2r^{\ast},\;{dr^{\ast}}/{dr}=1/{a(r)}$.
The smoothness of the metric at the junction $v=v_{0}$ requires
\begin{equation}
\frac{d\tilde{u}}{du}=\frac{\partial\tilde{u}}{\partial r}\frac{\partial
r}{\partial u}|_{v=v_{0}}=\frac{a_{0}\left(  \frac{2\ell^{2}}{u-v_{0}}\right)
}{a\left(  \frac{2\ell^{2}}{u-v_{0}}\right)  },
\end{equation}
and both regions can then be expressed in the $(u,v)$ coordinate system. The
corresponding spacetime structure is illustrated in Fig.~\ref{Fig3}. In addition, when
$\tilde{u}$ approaches infinity, the most probable value of $u$ is near
$v_{0}+2\ell^{2}/r_{h}$. Expanding $a(r)$ to first order around $r=r_{h}$, we have
$a(r)\simeq2\kappa(r-r_{h})$, and integrating the above equation yields
\begin{align}
\tilde{u} &  \simeq\int\frac{r_{h}^{2}}{2\ell^{2}\kappa}\left(  \frac
{2\ell^{2}}{u-v_{0}}-r_{h}\right)  ^{-1}\nonumber\\
&  \simeq-\frac{1}{\kappa}\ln\left[  12+\frac{6r_{h}(v_{0}-u)}{\ell^{2}%
}\right]  ,
\end{align}
and
\begin{equation}
u\simeq v_{0}+\frac{2\ell^{2}}{r_{h}}-\frac{\ell^{2}}{6r_{h}}e^{-\kappa
\tilde{u}}.\label{u}%
\end{equation}

Again, just like what has been done in the eternal case, we now denote the
coordinates of the island and the cutoff surface as $(v_{I},r_{I})$ and
$(v_{A},r_{A})$, respectively. The $u$-coordinates can then be derived 
using Eq.~(\ref{u}). Furthermore, a thermal bath coupled to radiation is
also required, and the conformal factor for the cutoff surface then is
different form that of the island. They are given by
\begin{equation}
\Omega_{A}^{2} =\frac{a_{0} \left(  \frac{2\ell^{2}}{u_{A}-v_{0}} \right)  }{
a\left(  \frac{2\ell^{2}}{u_{A}-v_{0}}\right)  } ,\quad\Omega_{I}^{2}
=a(r_{I}) \frac{a_{0} \left(  \frac{2\ell^{2}}{u_{I}-v_{0}} \right)  }{
a\left(  \frac{2\ell^{2}}{u_{I}-v_{0}}\right)  } .
\end{equation}

Considering the late stage of evaporation and assuming that the cutoff
surface is far away from the horizon, the generalized entropy takes the
following form
\begin{align}
S_{gen}  &  =f_{\mathbb{Q}}\frac{2\pi r_{I}^{2}}{G_{N}}+\frac{c}{12}\ln
[(v_{I}-v_{A})^{2}(u_{A}-u_{I})^{2}\Omega_{A}^{2}\Omega_{I}^{2}]\nonumber\\
&  \simeq f_{\mathbb{Q}}\frac{2\pi r_{I}^{2}}{G_{N}}+\frac{c}{12}\ln\left[
\frac{2a(r_{I})}{\kappa^{2}}(v_{A}-v_{I})^{2}\sinh^{2}\chi\right]  ,
\end{align}
where
\begin{equation}
\chi=\frac{\kappa}{2}(\tilde{u}_{I}-\tilde{u}_{A})=\frac{\kappa}{2}%
(v_{I}-v_{A}-2r_{I}^{\ast}+2r_{A}^{\ast}).
\end{equation}
The derivatives of $S_{gen}$ with respect to $v_{I}$ and $r_{I}$ are
\begin{align}
\frac{\partial S_{gen}}{\partial v_{I}}  &  =\frac{c}{12}\left(  \frac
{2}{v_{I}-v_{A}}+\kappa\coth\chi\right)  ,\\
\frac{\partial S_{gen}}{\partial r_{I}}  &  =f_{\mathbb{Q}}\frac{4\pi r_{I}%
}{G_{N}}+\frac{ca^{\prime}(r_{I})}{12a(r_{I})}-\frac{2c\kappa}{12a(r_{I}%
)}\coth\chi\nonumber\\
&  \simeq f_{\mathbb{Q}}\frac{4\pi r_{I}}{G_{N}}+\frac{c}{12a(r_{I}%
)}(a^{\prime}(r_{I})-2\kappa)-\frac{c\kappa}{3a(r_{I})}e^{-2\chi}\nonumber\\
&  \simeq f_{\mathbb{Q}}\frac{4\pi r_{I}}{G_{N}}-\frac{c\kappa}{3a(r_{I}%
)}e^{-2\chi}.
\end{align}
The above equations can be rewritten in a more compact form,
\begin{align}
v_{A}-v_{I}  &  =\frac{2}{\kappa\coth\chi},\\
a(r_{I})  &  =\frac{1}{f_{\mathbb{Q}}}\frac{cG_{N}\kappa}{12\pi r_{I}%
}e^{-2\chi}.
\end{align}
Without the need of exact island's coordinates, the final expression of
radiation entropy becomes
\begin{align}
S_{R}  &  =f_{\mathbb{Q}}\frac{2\pi r_{h}^{2}}{G_{N}}+\frac{c}{12}\ln\left[
\frac{1}{f_{\mathbb{Q}}}\frac{2cG_{N}}{3\pi r_{h}\kappa^{3}\coth^{2}\chi
}e^{-2\chi}\sinh^{2}\chi\right] \nonumber\\
&  \simeq f_{\mathbb{Q}}\frac{2\pi r_{h}^{2}}{G_{N}}+\frac{c}{12}\ln\left[
\frac{1}{f_{\mathbb{Q}}}\frac{2cG_{N}}{3\pi r_{h}\kappa^{3}}\right]  .
\end{align}
This result possesses the following advantages:

\begin{itemize}
\item[(1)] At late times in the black hole evaporation process, the radiation
entropy becomes time-independent and saturates at a finite value. The
disappearance of the cutoff-surface dependence implies that the radiation
entropy no longer diverges as the cutoff is moved farther away. One might
wonder that, as the cutoff surface recedes, the distance between the quantum
extremal surface and the cutoff surface increases, and according to the
formula of $s$-wave approximation, one would expect a logarithmic divergence
(under an eternal black hole, such a divergence would scale with the cutoff's
position rather than its logarithm). In practice, however, this is not the
case. Under the $s$-wave approximation, what is calculated is the spacetime
interval between the two surfaces. A larger spatial separation does not
necessarily lead to an increase in the spacetime interval, as this also
depends on the temporal separation. It is precisely this subtle interplay that
prevents the radiation entropy from diverging as the cutoff surface is moved
farther away.

\item[(2)] The radiation entropy depends solely on the properties of the
horizon and other constant parameters, such as the $f(\mathbb{Q})$ model,
central charge and gravitational constant. This outcome not only agrees with
the general properties of entanglement entropy but also aligns with the
predictions of quantum gravity
\cite{Fursaev:1994te,Kaul:2000kf,Page:2004xp,Zhang:2008rd,Ghosh:2004rq,Solodukhin:1997yy}%
, in which the black hole entropy has a logarithmic area correction.

To see this point more transparently, we can rewrite the radiation entropy as
\begin{equation}
S_{R}=\frac{f_{\mathbb{Q}}\mathcal{A}_{H}}{2G_{N}}-\frac{c}{6}\ln
\frac{\mathcal{A}_{H}}{4G_{N}}+\frac{c}{12}\ln\frac{64c\ell^{2}}{9\pi
f_{\mathbb{Q}}},
\end{equation}
where we have used $\kappa={3r_{h}}/{(2\ell^{2})}$ and the properties of
logarithmic function. If the coefficient in front of the geometric term can be
measured with sufficient accuracy, it may provide valuable insights into the
functional form of the $f(\mathbb{Q})$ gravity. For instance, if the geometric
contribution preserves its original form, this would imply $f_{\mathbb{Q}%
}=1/2$. Compared with constraining the the $f(\mathbb{Q})$ function through cosmological
observations, the constraint from the information loss problem arises directly
from theoretical self-consistency.
\end{itemize}

\subsection{Page Time}

To determine the Page time, we first need to evaluate the radiation entropy in
the early stage of black hole evaporation, when the island is generally not
assumed to have formed. For a collapsing black hole, this evaluation can be
implemented by placing the island in a pure AdS background and taking its
spatial coordinate $r$ to approach zero. Using the island formula, the
early-time radiation entropy is given by
\begin{align}
S_{R}  &  =\lim_{r_{I}\rightarrow0}\frac{c}{12}\ln\left[  (u_{I}-v_{A}%
)^{2}(v_{A}-v_{I})^{2}a_{0}(r_{I})\frac{a_{0}\left(  \frac{2\ell^{2}}%
{u_{A}-v_{0}}\right)  }{a\left(  \frac{2\ell^{2}}{u_{A}-v_{0}}\right)
}\right] \nonumber\\
&  =\frac{c}{12}\ln\left[  \frac{12r_{h}}{\ell^{2}}(v_{A}-v_{I})^{2}e^{\kappa
v_{A}}\right] \nonumber\\
&  \propto\frac{c\kappa v_{A}}{12}\quad(v_{A}\rightarrow\infty).
\end{align}
It is easy to see that in the early stage of evaporation, the radiation
entropy increases with time, which is due to the accumulation of interior
Hawking quanta generated by the outgoing radiation. At late times, however,
this island-free radiation entropy diverges, indicating that the contribution
from the island must be taken into account.

The Page time can subsequently be obtained by equating the island-free
radiation entropy with that of the island contribution. By keeping only the
leading-order terms, we obtain
\begin{equation}
v_{Page}=f_{\mathbb{Q}}\frac{16\pi\ell^{2}r_{h}}{cG_{N}}.
\end{equation}
It can be seen that the Page time is proportional to the black hole horizon
radius and is also affected by the choice of the $f(\mathbb{Q})$ model.

\section{Conclusion}
\label{secIV}

In this work, we investigate the radiation entropy in the framework of
$f(\mathbb{Q})$ gravity, focusing on a class of asymptotically AdS black
holes. Specifically, by computing the thermodynamic entropy, we find that the coefficient
of the area term in the entropy must be modified, thereby encoding the information about the
underlying gravitational model. According to the origin of
the island rule in AdS/CFT correspondence, this further implies that the geometric contribution
associated with the quantum extremal surface should be modified in the same
manner when evaluating the radiation entropy. For the eternal black hole,
although the island rule yields a time-independent radiation entropy, it diverges as the
cutoff surface is moved far away. This finding indicates that even with a coupled
thermal bath, the s-wave approximation remains invalid due to the presence of
ingoing modes. Such an issue is avoided in the collapsing black hole
background. By coupling the system to a thermal bath and applying the island rule, we
obtain a logarithmic correction to the area law in the radiation entropy,
which is consistent both with the general properties of entanglement entropy
and with the predictions of quantum gravity theories. Furthermore, since the radiation
entropy and Page time explicitly contain information about the $f(\mathbb{Q})$ gravity,
our results provide a new theoretical avenue for constraining the functional
form of $f(\mathbb{Q})$.

\bigskip

\acknowledgments

This work is supported by National Natural Science Foundation of China (NSFC)
with Grant No. 12375057 and the Fundamental Research Funds for the Central
Universities, China University of Geosciences (Wuhan).

\end{document}